# Logic Verification of Ultra-Deep Pipelined Beyond-CMOS Technologies

Arash Fayyazi, Shahin Nazarian, and Massoud Pedram
Ming Hsieh Department of Electrical and Computer Engineering, University of Southern California
{fayyazi, shahin.nazarian, pedram}@usc.edu

*Abstract*—Traditional logical equivalence checking (LEC) which plays a major role in entire chip design process faces challenges of meeting the requirements demanded by the many emerging technologies that are based on logic models different from standard complementary metal oxide semiconductor (CMOS). In this paper, we propose a LEC framework to be employed in the verification process of beyond-CMOS circuits. Our LEC framework is compatible with existing CMOS technologies, but, also able to check features and capabilities that are unique to beyond-CMOS technologies. For instance, the performance of some emerging technologies benefits from ultra-deep pipelining and verification of such circuits requires new models and algorithms. We, therefore, present the *Multi-Cycle Input Dependency (MCID)* circuit model which is a novel model representation of design to explicitly capture the dependency of primary outputs of the circuit on sequences of internal signals and inputs. Embedding the proposed circuit model and several structural checking modules, the process of verification can be independent of the underlying technology and signaling. We benchmark the proposed framework on post-synthesis rapid single-flux-quantum (RSFQ) netlists. Results show a comparative verification time of RSFQ circuit benchmark including 32-bit Kogge-Stone adder, 16-bit integer divider, and ISCAS'85 circuits with respect to ABC tool for similar CMOS circuits.

## I. INTRODUCTION

The ongoing demand for energy-efficient and high-performance computing has driven the development of semiconductors since its early days, but with the conclusive end of Moore's Law and rising challenges to the physical scaling of CMOS devices [1], there is a significant need for new device technologies to continue beyond end-of-scaling CMOS technology. The exploration and study of novel logic components has been a main research focus in the past decade [2], in pursuit of extending the semiconductor industry roadmap beyond the CMOS technology [2]. Beyond-CMOS device concepts include a wide variety of elements such as charged-based components like Quantum-dot Cellular Automata (QCA) [3]. Additionally, the research community has also focused on exploring non-charge-based solutions such as spin-based components like Spin Wave Devices (SWD) [4] and NanoMagnetic Logic (NML) [5],[6]. Superconducting technologies such as rapid single-flux-quantum (RSFQ) [7], the quantum flux parametron (QFP) [8], reciprocal quantum logic [9], energy-efficient RSFQ (eSFQ) [10], and Adiabatic QFP (AQFP) [9] are also very promising candidates, given their potential to be 1000x as energy efficient as the state-of-the-art CMOS technologies Such high levels of energy efficiency are strongly required for high-performance computers having performances in exa-FLOPS. As an example power consumption of CMOS technologies could exceed 100 MW, which is equal to the power generated by a small power plant [11]. Beyond-CMOS technologies, however have several constraints that prevents them from supporting complex designs. One of the constraints in these technologies is that in order to cascade elementary devices, the complete circuits need to be clocked [9], [12], [13]. Therefore, state of computer-aided design (CAD) tools available to these technologies communities has been both outdated and not suitable for complex designs.

Logical equivalence check (LEC) is one of the most important checks during the entire chip design process [14]. Design passes through various steps like synthesis, ECOs (engineering change orders), and numerous optimizations, therefore it is vital to efficiently verify that the logical functionality remains intact and does not break because of any of the automated or manual changes. Hence, developing suitable, LEC techniques for beyond-CMOS devices will reduce verification time and ensure the correctness of the circuit functionality as the complexity of circuits grows [14]. In this paper we introduce a novel logical equivalence check (LEC) framework for beyond-CMOS circuits. The proposed LEC framework is technology independent, i.e., it is able to verify not only the CMOS technologies, but also various candidate technologies of future thanks to its independence to details of technology such as timing and signaling requirements.

We summarize our contributions as follows:

- We propose a novel graph representation of the beyond CMOS circuit, which we refer to as the *multi-cycle input dependency (MCID)* circuit model. MCID represents functional behavior model of a clock-synchronous pipelined netlist and explicitly captures the dependency of primary outputs of the circuit on sequences of internal signals and inputs which affect outputs. This representation model enables unifying timing and functionality pieces of

The research is based upon work supported by the Office of the Director of National Intelligence (ODNI), Intelligence Advanced Research Projects Activity (IARPA), via the U.S. Army Research Office grant W911NF-17-1-0120. The views and conclusions contained herein are those of the authors and should not be interpreted as necessarily representing the official policies or endorsements, either expressed or implied, of the ODNI, IARPA, or the U.S. Government. The U.S. Government is authorized to reproduce and distribute reprints for Governmental purposes notwithstanding any copyright notation herein. This project is also supported in part by a grant from the Software and Hardware Foundations program of the National Science Foundation.
Authors are with University of Southern California (USC).

information of the given circuit into a unique functional model. Mitering a MCID model and golden model, correctness of functionality of the given circuit can be verified by a customized LEC which uses *Boolean Satisfiability* (SAT) as an underlying reasoning engine.

- We also present an input timing control logic (ITCL) which handles different arrival times of inputs at primary inputs since inputs can be applied at different clock periods.

- We propose two algorithms for structural checks on 1) pin count nets (e.g., typically two in SFQ technologies) and 2) equalization of path delay (to ensure coherent data wave propagation).

- We have also implemented a parameterized LEC tool to validate the correctness of our approach in verifying the beyond-CMOS circuit functionality independently of the underlying technology details including signaling and timing. Depending on technology, parameters are set to abstract the timing and signaling information. For example, our tool is able to perform LEC on SFQ vs CMOS, or SFQ vs AQFP. With respect to the LEC in ABC [15], as a baseline, our tool obtains comparative verification time on a set of benchmarks including 32-bit Kogge-Stone adder (KSA), 16-bit integer divider and ISCAS'85 circuits.

The remainder of this paper is organized as follows. In Section II, we introduce the central concepts used in this work. Section III presents the required algorithms and models for LEC of beyond-CMOS technologies. Section IV presents benchmarking results for our proposed LEC methodology, followed by conclusions in Section V.

## II. BACKGROUND & MOTIVATION

Hereafter, we introduce the general operating principles of beyond-CMOS technologies.

### A. Ultra-Deep Pipelining

Ordinary pipelined systems [16], [17] can process more than one instructions on a set of data simultaneously and are divided to several stages, isolated by registers. Each of these stages nominally performs its part an operation (i.e., instruction) separately from rest of the stages. The data flow through each stage is determined by the global clock signal which allows processing of a new set of data only once the previous set has propagated to the next stage.

In contrast, ultra-deep pipelining may utilize each logic cell (gate) as one stage. This is feasible since beyond-CMOS gates may need a clock signal to operate. For instance, SFQ gates (except for non-clocked gates such as confluence buffer, Splitter, I/O cells, and T-Flip-Flops) need a clock signal to function. As an example, Fig. 1 shows the circuit diagram of an SFQ OR gate and the corresponding waveform to show its functionality. After the arrival of an input pulse, the arrival of clock pulse, results in an output pulse. This is interpreted as logic 1. However, no input pulse, results in no output pulse and this would be interpreted as logic 0.

Similar operating principles may exist for the variety of beyond-CMOS technologies including *SWD* [4], *QCA* [3], and *NML* [5], [6]. In [18], the authors proposed an efficient synthesis framework for aforementioned technologies. Zografos et al. [18] suggested that to take advantage of the non-volatility property (which would eliminate the need for a constant supply voltage and reducing the standby power consumption), all gates in the circuit need to be clocked in order to cascade elementary devices [4], [12].

In order to make use of both logic and memory capabilities of these beyond-CMOS technologies, their framework tackles the physical constraints of the circuits based on these devices by equalization of path delay to ensure coherent data wave propagation which is equivalent to ultra-deep pipelining introduced here.

For a gate in such technologies to operate correctly, all of its fanin gates should have the same logic level. If there is a difference among logic levels of fanins of a gate, some D Flip-

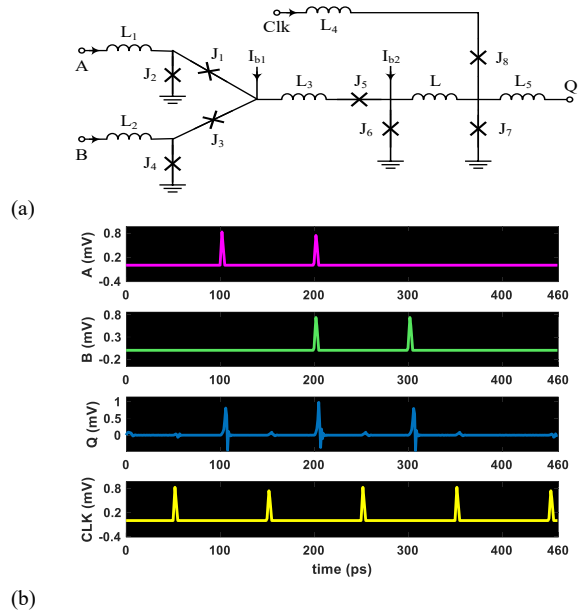

Fig. 1. SFQ OR gate (a) equivalence circuit (b) corresponding signal waveform.

Flops (DFF, or buffer, dependent on the underlying technology) should be inserted into outputs of fanin gates with smaller logic levels. For example, as shown in Fig. 2 if the first fanin (in1) of an AND2 gate has a logic level of one and the second fanin (in2) has a logic level of zero, one DFF should be added to the input of in2. Without path-balancing, correct pulses on in2 will be consumed by this AND2 gate one clock before arrival of the corresponding pulses on the first input, hence, this gate will not be able to produce correct output values. One of the most important constraints which ultra-deep pipelining imposes is that all the propagation paths from the combinational circuit's inputs to outputs have approximately the same logic level (i.e., number of gates from primary inputs), then each data flow propagates uniformly to the outputs without interfering with adjacent flow.

### B. Fanout Restriction

Synthesis tools for beyond-CMOS implementation should normally limit the cascading of one component, ensuring feasibility, given that several emerging technologies have no intrinsic gains [9], [13], [18]. For instance, in SFQ logic family, if a gate needs to have more than one fanout, a special SFQ gate called Splitter should be added to the output of this gate. Splitter is an asynchronous gate that accepts an SFQ pulse and produces an output pulse on each of its fanouts after its intrinsic delay. One Splitter can produce only two fanouts. For additional fanouts, more Splitter cells should be added in a binary tree structure. To have n fanouts, n−1 Splitter cells are needed.

Similarly, Zografos et al. [18] confirmed that one of the physical constraints of the circuits based on SWD [4], QCA [3], NML [5], [6] is fanout restriction that needs to be addressed so that the resulting circuit can be efficiently implemented in the selected technologies. Fanout limitations for different technologies may vary. For instance, for AQFP, Splitter cells are clocked buffers that can have 1-to-2, 1-to-3, and even 1-to-4 fanouts [9].

## III. OUR VERIFICATION FRAMEWORK

This section introduces our LEC framework (see Fig. 3) along with the specific assumptions for the selected technologies. Additionally, the MCID graph modeling of beyond-CMOS circuits is described. Our framework takes into account the fundamental differences between beyond-CMOS and CMOS circuits. It starts with structural checks and then builds the MCID model. Finally, it verifies the correct functionality of the given netlist using customized LEC functions.

### A. Structural Checkers

The proposed LEC framework extracts the circuit network of gates and wires from gate-level structural model and analyzes the network to ensure that the circuit satisfies the fanout restriction (e.g., single fanout in the case of SFQ gates except for Splitter cells which can have a fanout of two) and also meets the path-balanced requirements to needed for ultra-deep pipelining. To verify these properties, we utilize two checkers, a circuit fanout checker followed by a circuit path-balancing checker.

First, we use the following definitions provided in [18]:

- Distance ($D$) between two different components, is the set of lengths of any path going from the source to the destination.

- Base distance ($BD$) of a component, is the set of lengths of any path going from any netlist input to that component. The maximum length in this set represents the *depth* of the component.

The fanout checker ensures that the given netlist satisfies the fanout restriction. For instance, in case of SFQ technology, Splitter cells must have been inserted to adjust the SFQ gate fanout for any logic cell driving two gates or more. The fanout checker extracts the circuit's wire adjacency lists and verify that the size of each list is not larger than 1 (or 2 for Splitter cells).

Objectives of the path balancing checker are as follows, (a) all paths from one component to another must be equal in length; (b) maximum base distance of all netlist outputs must be equal. Differently worded, for any two connected components the minimum distance must be equal to its maximum distance. If the first goal is also obtained, then the base distance of all outputs must be equal as this set will only contain one number (see

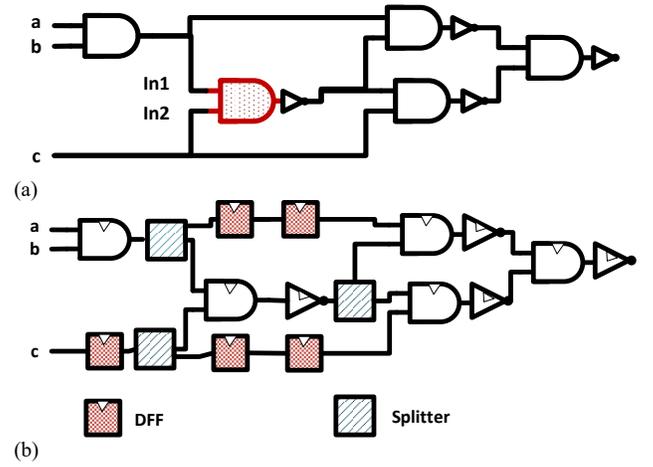

Fig. 2. (a) A digital CMOS circuit (b) Counterpart SFQ circuit where DFFs and Splitter cells are added for path-balancing and Fanout limitation, respectively.

Theorem 1). Hence, the path balancing checker (see Algorithm 1) is a customized Depth First Search (DFS) that assumes that the circuit is represented as a Directed Acyclic Graph (DAG). Thus, the customized DFS is constrained to iterate over just the primary inputs of the circuit. The recursive DFS visits have been customized, which called *qPathDepthCounter*, to handle Splitter cells as they do not increase logic level or path depth but present two separate paths to recurse on. Both the fanout and path balancing checkers have worst-case runtime complexity of $O(|G|)$, where $|G|$ is the number of gates in the circuit. If the first checker is failed, the given circuit is flagged for the corresponding fanout error and rejected from undergoing verification in LEC framework (cf. Fig. 3), thus saving verification time.

**Theorem 1.** If the base distance of all outputs is equal as this set will only contain one number, then all paths from one component to another must be equal length.

**Proof.** Proof by contrapositive. The contrapositive of the above statement is, if two paths ($P_1$ and $P_2$) from one component ($comp_1$) to another ($comp_2$) are not equal in length, the base distance of at least one output ($PO_y$) contains at least two numbers. Since there is at least one path from one primary input ($PI_x$) to $comp_1$ (e.g., $P_{PI_x,1}$) and similarly from $comp_2$ to one primary output ($P_{2,PO_y}$) there exist two unequal-length paths from $PI_x$ to $PO_y$ which are $\{P_{PI_x,1}, P_1, P_{2,PO_y}\}$ and $\{P_{PI_x,1}, P_1, P_{2,PO_y}\}$. So, Theorem 1 is proved.

### B. MCID and ITCL

Equal propagation paths from inputs to outputs is a sufficient, but not necessary condition. More specifically, we have observed that an SFQ circuit may not be fully path-balanced yet

---

**Algorithm 1.** Path-Balancing Checker

**Input:** PI (circuit's Primary input list), G (circuit's DAG)
**Output:** Path balancing checker result
1:   **For** each element in PI **do**
2:   |  *BaseDistanceList = BaseDistanceList ∪ qPathDepthCounter()*
3:   **If** *BaseDistanceList* values are not all one value **then**
4:   |  **Return** Fail
5:   **Return** Pass

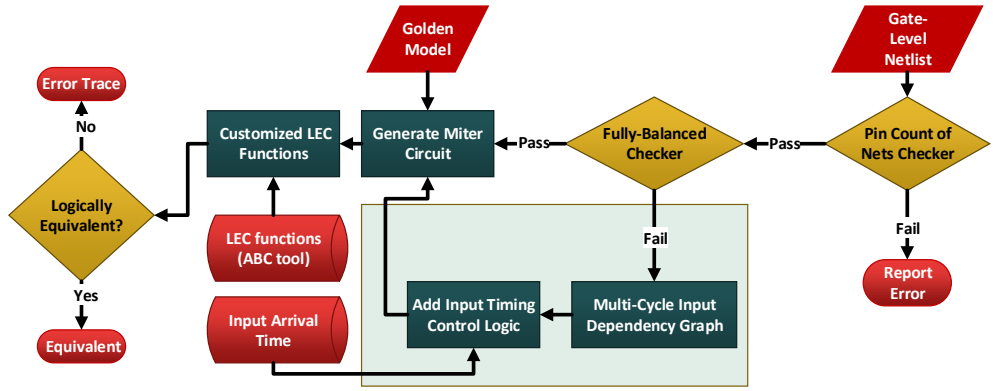

Fig. 3. Flowchart of proposed LEC framework.

function correctly. A high-throughput Arithmetic Logic Unit (ALU) SFQ design with different arrival time of inputs [19], [20], custom design circuits with false paths [21], and synthesis methodology of area-efficient SFQ circuit [22]. Therefore, structural checkers are not sufficient to check specific properties of the new technologies. Hence, the following approaches are introduced and implemented as part of the proposed LEC.

*1) MCID Graph*

Functional error because of partially-balanced paths only can be captured by observing circuit behavior during several clock cycles. This is due to fact that the output result of partially-balanced circuit depends on the inputs of multiple previous clock cycles not only one specific clock cycle which is the case in CMOS technology. In the following we introduce several definitions that will be used in this section.

- Each clocked gate generates a pulse at its output in the following clock cycle after receiving required pulses at its input(s). Hence, delay of each gate is assumed to be one clock cycle as also referred to as one *time unit*.

- *Path delay* is defined by the number of time units required by a signal to propagate on that path which is equal to number of clocked gates.

In a circuit where the shortest path delay from inputs is $D_{p_s}$ time units, and the longest path delay is $D_{p_l}$ time units, the current output $O_t$ depends on the inputs of $I_{t-D_{p_s}}, I_{t-D_{p_s}-1}, \ldots, I_{t-D_{p_l}}$ where indices denote the times of input with a step of one time unit. Each index is called a time step (i.e., time unit). Upper bound of number of time units to observe is circuit longest path. In simpler words, the older inputs (the inputs of the time units more distant from current time $t$) influence the output through longer paths and the newer inputs (the inputs of the time units closer to the current time $t$) affect the output through shorter paths. LEC tools for CMOS circuits check only the functionality of circuits over all possible input vectors at one specific time step which is meaningless in the case of beyond-CMOS circuits (see Fig. 4) Hence, we propose a clock-synchronous logical model of a circuit explicitly capturing multi-cycle input dependencies which helps to unify time and functional pieces of information into the functional domain.

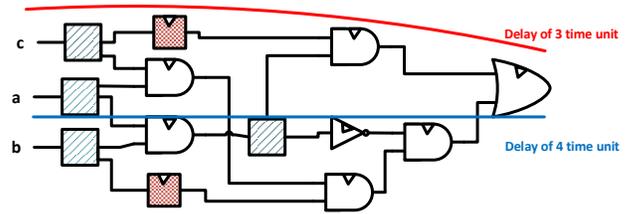

Fig. 4. An example of SFQ circuits with different path delays from inputs to outputs.

The MCID model represents the functional behavior model of a clock-synchronous pipelined netlist which explicitly captures the dependency of primary outputs of the circuit on sequences of inputs which affect outputs. This model enables analyzing the functional behavior of beyond-CMOS circuits using customized LEC functions. The underlying idea is that a signal $s_t$ represents a signal s of the given circuit at time step $t$. The process of constrcuting MCID graph is described in Algorithm 2. The input data of this algorithm is a circuit netlist. For each gate the algorithm creates as many copies of values of the gate at different time steps as necessary to determine the output. The algorithm starts from Primary Outputs (PO) and traverses the given circuit graph backward (line 2). Given the delay of one time unit to each clocked gate in the given circuit, the output of a gate depends on its corresponding inputs one time step before. These inputs are given by the predecessor node of the output in the given circuit graph (line 6). Given the output and its driving inputs, a new gate is created in which the output and the inputs have the timing difference of one time unit (lines 11 and 13). Note that non-clocked gates like Splitter cells must be treated differently. Such gates don't receive any clock signal, so the output (e.g., pulse in RSFQ technology) of them is generated at the same time step as input of them. We copied the non-clocked gate (except Splitter whose functionality is the same as buffer and we can skip it) and also find its predecessor node at the same time step (line 7-11). The inputs of gates at the current step are collected in the set *SIG_temp* to be used for the next backward traversal step (lines 14–16). If the input is a *Primary Input* (PI) or already exists in the set *SIG_temp* (fanout case), or if it will not be added to the set *SIG_temp* (line 15). This new circuit called the MCID circuit. MCID model generation has worst-case runtime complexity of $O(|G|)$.

**Theorem 2.** Assuming $MCID_t$ as the MCID model for the the PO at time step t ($PO_t$), it is guaranteed that $MCID_t$ models all time steps of internal signals and inputs which affect $PO_t$.

**Proof.** Proof by construction. The MCID model is constructed by traversing a given circuit one step time wise via *Breadth-First Search* (BFS) from the output PO towards PIs. Algorithm 2 visits every gate at least once. If a gate g is visited, three cases may occur as follows:

- Case 1: The gate g is visited only once. In this case, signal s of the output of gate g never reconverges. Because if signal s reconverges, gate g is visited at least twice while traversing the circuit by the algorithm from the output PO backwards. Therefore, the value of signal s only at one time step affects the value of $PO_t$.
- Case 2: The gate g is visited through paths with the same *path delays*. In this case, only the value of signal s at one time step affects output o through multiple paths. Because the paths have the same *path delays* to output o. Line 15 in Algorithm 2 checks whether a gate has been visited through paths with the same *path delays*. If a gate is visited for the first time through a path, the *if*-expression becomes true. But if a gate is visited for the second time through a path with the same *path delay*, the *if*-expression becomes false and the corresponding gate only once at one time step is copied.
- Case 3: Gate g is visited through paths with different *path delays*. In this case, the value of signal s at different time steps affects $PO_t$. In this case, the *if*-expression of line 15 in Algorithm 2 is false every time. Therefore, gate g is copied for all corresponding time steps.

Hence, the $MCID_t$ constructed by Algorithm 2 models all time steps which may affect $PO_t$. This proves Theorem 2.

---

**Algorithm 2.** MCID (Multi-Cycle Input Dependency) Constructor

**Input:** Circuit netlist, PI (circuit's primary input set), PO (circuit's primary output set), *NCG (non-clocked gates set)*
**Output:** MCID model
1:  *Clock_cycle* = 0
2:  *SIG* = *PO*
3:  **While** *SIG* ≠ { } **do**
4:      *SIG_temp* = { }
5:      **For** each *sig* ∈ *SIG* **do**
6:          *gate* = *predecessor(sig)*
7:          **If** *gate* ∈ *NCG* **then**
8:              **If** *gate* != *Splitter* **then**
9:                  | *MCIDModel.copy* (*gate*, $i_{t-Clock\_cycle}$, $o_{t-Clock\_cycle}$)
10:             *gate* = *predecessor(sig)*
11:             *MCIDModel.copy* (*gate*, $i_{t-Clock\_cycle-1}$, $o_{t-Clock\_cycle}$)
12:         **Else**
13:             | *MCIDModel.copy* (*gate*, $i_{t-Clock\_cycle-1}$, $o_{t-Clock\_cycle}$)
14:         **For** each *input* ∈ *I(gate)* **do**
15:             **If** *input* ∉ *SIG_temp* and *input* ∉ *PI* **then**
16:                 | *SIG_temp* = *SIG_temp* ∪ *input*
17:     *SIG* = *SIG_temp*
18:     *Clock_cycle++*
19: **Return** *MCID*

---

*2) ITCL*

Input timing control logic unit (ITCL) is added for controlling the inputs arrival time. Assume that a circuit is designed such that its inputs pulses must be applied in different clock cycles. To handle such cases, *ITCL* adds buffer(s) gate to MCID circuit inputs which receive input pulses later than other(s). For instance, the inputs arrival time given by designer is

$$T(I): [t_{I_1}, t_{I_2}, \ldots, t_{I_n}] \qquad (1)$$

where $t_{I_x}$ represents the arrival time of input $I_x$. ITCL finds $\underset{x}{argmin}\{t_{I_x} | I_x \in I\}$, i.e., *minI*, and adds $t_{I_x} - t_{I_{minI}}$ buffer(s) to input pin *x* of MCID circuit.

To generate the miter circuit of combining MCID model and golden model, the common inputs must be matched. Note that the main key in customized LEC functions is the use of structural similarities to reduce the size of SAT instance and hence to speed-up the LEC process. However, in the final MCID circuit, we may have several inputs in different clock cycle that can be matched to one primary input of golden model. So, to help customize the LEC function for finding structural similarities, it is crucial to find the largest subset of inputs arrived at same clock cycle of MCID circuit and match them to the primary inputs of golden model. Hence, there may be input(s) of MCID model that does not connected to any input of golden model. Overall, the final generated miter circuit is shown in Fig. 5. Following are the details of miter circuit,

$$\varphi_1 = C_1(I_1(t)), \varphi_2 = C_2(I_2) \qquad (2)$$

Fig. 6 shows an example of building MCID circuits and then applying ITCL. The SFQ circuit in Fig. 6 (a) is supposed implement $\bar{a}bcd$ but it is not fully path-balanced. So, using the corresponding MCID circuit (see Fig. 6 (b)) an error trace can be found, i.e.,

$I_1(t) = \{[a = 0, b = 1, c = 1, d = 1]_{clock\ cycle\ =\ 0},$
$\qquad\qquad [a = 0, b = 1, c = 1, d = 0]_{clock\ cycle\ =\ 1}\}$
$I_2 = [a = 0, b = 1, c = 1, d = 0]$
$\varphi_1 = C_1(I_1(t)) = 1 \neq \varphi_2 = C_2(I_2) = 0$

Note that MCID model is in functional domain, so, only functionality of all gate is considered in customized LEC functions. That's the reason for removing clock pin from each gate and replacing the DFF with buffers. However, if the designer makes the circuit partially-balanced on purpose since input d arrives one clock cycle later than other inputs, ITCL adds one buffer in front of that input pin as shown in Fig. 6 (c) and the circuit meets its specification.

## IV. EXPERIMENTAL RESULTS

In this section, the efficacy of the proposed LEC is investigated. First, we introduce SFQ technology which we used for benchmarking our LEC framework. Next, the experimental setup and experimented combinational SFQ circuits are described. Two types of design errors, namely functional and structural errors are inserted to the netlist. Finally, the performance of MCID model in detecting these errors is presented.

## A. Technology

Developed in the late 1980s, a very promising family of "beyond-CMOS" devices are single flux quantum (SFQ) circuits based on Josephson junction [7]. Similar to how CMOS circuits are built with transistors (3-terminal devices) as their active elements, SFQ circuits are built using Josephson junctions (JJ, 2-terminal devices) as their active components. When

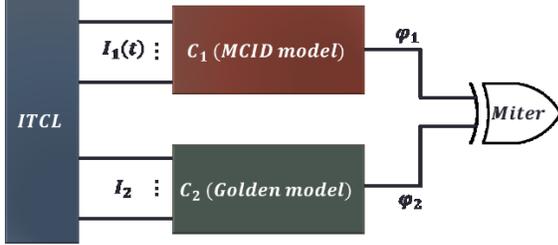

Fig. 5. Miter Circuit.

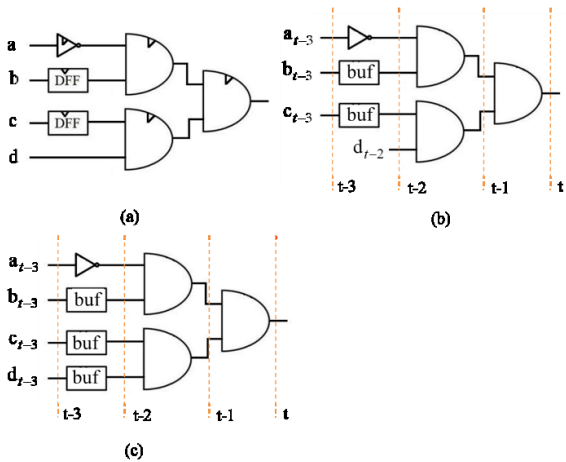

Fig. 6. Running Example of building MCID circuit and applying ICTL (a) An SFQ circuit (b) corresponding MCID circuit (c) final circuit after applying ICTL with considering that input $d$ arrives one clock cycle later than other inputs.

Josephson Junctions are operated at cryogenic temperatures, these superconducting devices exhibit the Josephson effect - a phenomenon of a current called super-current that flows indefinitely long without any applied voltage. As a result, SFQ circuits benefit from Josephson junctions with high switching speeds on the order of picoseconds and low switching energy on the order of $10^{-19}$ joules at 4.2 Kelvin [23]. The switching energy of Josephson junctions is two to three magnitudes lower than that of end-of-scaling CMOS devices [1]. Thus, SFQ circuits have demonstrated the potential to achieve the computing demands for energy-efficient and high-performance circuits [10], [24]. The rapid SFQ (RSFQ) technology is a new version of SFQ in which the parameter margins of the SFQ as well as operation speed to 300 GHz are increased [7]. (R)SFQ gates are pulsed-based and the presence and absence of a pulse are considered as "1" and "0", respectively. A pulse is a single quantum of magnetic flux ($\Phi_0 = h/2e = 2.07mV \times ps$) with a duration of a few ps and amplitude of a few mV.

## B. Experimental Setup

The computer system used for the testing utilized an Intel Core i7-7700HQ CPU with nominal clock frequency of 2.8 gigahertz and 16 gigabytes of RAM. Among the available (open source) tools, ABC [15] was selected as the baseline verification tool. ABC is relatively fast and scalable compared to other open-source tools. The considered combinational SFQ circuits include Kogge-Stone adders (KSA), array multipliers, integer dividers, and ISCAS'85 combinational benchmark circuits [25]. Since our tool objective is to verify correct functionality of post-synthesis netlist, we used [13] as a synthesis tool to generate equivalent gate-level structural model from given a combinational circuit. Parameters including number of fanout constraints, non-clocked gates, and an indicator for ultra-deep pipelined circuits are set to abstract the timing and signaling information.

In order to evaluate the effectiveness of MCID model, two types of design errors were inserted to the netlist, functional and structural errors. Functional errors are those which change the functionality of circuit by swapping gates. As post synthesis SFQ circuits should be fully path-balanced and each gate should have limited fanout count, there exists two kinds of structural errors. One is to make the circuit partially-balanced by deleting a DFF. The other is to increase the number of fanouts of a gate by removing a Splitter.

## C. LEC Results

With these inserted errors, the proposed LEC framework was evaluated under circuit designs to indicate the effectiveness of the proposed checkers and models, specifically in terms of verification scalability and runtime.

### 1) Performance of Proposed LEC

Each entry of runtime and number of gates after applying MCID model are average of ten runs of LEC framework for different type of errors. The functional errors were generated by swapping different kinds of gates. The Structural errors were created by randomly removing one or two DFFs near the primary outputs and by randomly removing one Splitter. Verification results are summarized in Table I. The results for 4-, 8-, 16-bit array multiplier as well as integer divider and KSA have been derived but are not included in this paper for brevity. One of the advantages of implemented tool is that it can generate the error trace similarly to conventional LEC tools. Furthermore, it also keeps track of the information about input vector arrival time. In other words, it has information about values of each input at different clock cycles that generate the wrong output. The results show that the proposed framework detects functional and structural errors in a very short runtime. Meanwhile, for partially-balanced circuits, it also shows the number of gates in the MCID model. We believe the following are the reasons for the short runtime of the proposed LEC, the used benchmark [25] consists of combinational logic circuits. The Verilog implementations of those circuits are almost all data-flow style. In data-flow style, the behavior of the logic is represented by Boolean operations and assignments. The corresponding networks of Verilog modules are therefore very similar to those of the synthesized netlists. As the proposed LEC utilizes the structural hashing (as the ABC does) of the internal nodes, the equivalence checking process of those similar (pre-synthesis and post-synthesis) structures would be fast. To validate this, we assess the implemented tool under different types of an 8-bit multiplier and report the results in section IV.C.3.

TABLE I. THE PROPOSED LEC PERFORMANCE.

| Benchmark | Type of error | Run Time | # of Gates | | |
|---|---|---|---|---|---|
| | | | Golden | SFQ | MCID circuit |
| c432 | No error | 0.01 | 209 | 1699 | N/A |
| | Functional | 0 | 209 | 1699 | N/A |
| | Structural | 0 | 209 | 1698 | 2808 |
| c499 | No error | 0.01 | 202 | 868 | N/A |
| | Functional | 0.01 | 202 | 868 | N/A |
| | Structural | 0.007 | 202 | 866 | 694 |
| c880 | No error | 0.04 | 357 | 1427 | N/A |
| | Functional | 0.01 | 357 | 1427 | N/A |
| | Structural | 0 | 357 | 1425 | 1571 |
| c1355 | No error | 0.04 | 514 | 868 | N/A |
| | Functional | 0 | 514 | 868 | N/A |
| | Structural | 0.008 | 514 | 867 | 692 |
| c1908 | No error | 0.04 | 880 | 1474 | N/A |
| | Functional | 0.01 | 880 | 1474 | N/A |
| | Structural | 0 | 880 | 1472 | 1556 |
| c3540 | No error | 0.13 | 1667 | 3485 | N/A |
| | Functional | 0.06 | 1667 | 3485 | N/A |
| | Structural | 0.03 | 1667 | 3483 | 4539 |
| c6288 | No error | 0.46 | 2416 | 6189 | N/A |
| | Functional | 0 | 2416 | 6189 | N/A |
| | Structural | 0.02 | 2416 | 6188 | 8366 |

*2) MCID Model Size*

An example of duplicating a gate is showed in Fig. 7. By inserting a structural error (i.e., removing the DFF in this example), the circuit becomes partially-balanced and INV 1 is duplicated in MCID model.

As discussed in Section III.B, gates get copied when netlists have outputs with different base distances. So, in order to find the upper bound of MCID circuit size, closest DFFs to POs should be removed and there must be at least one Splitter before them to create paths with different delays. In the MCID model, gates before Splitter will be duplicated. Since the fanout count is limited, the MCID model is *linearly* larger than the original circuit. However, the number of gates still can exponentially grow based on the logical level of the closest removed DFF to PO and number of reconvergent paths through removed DFFs. The upper bound number of gates that are duplicated in MCID for one primary output in the case one reconvergent path is,

$$\# \text{ of added gates} = 2^{D_{p_{Spl}}-1} + \cdots + 2^0 = 2^{D_{p_{Spl}}} - 1 \quad (3)$$

where $D_{p_{Spl}}$ is the path delay of Splitter (i.e., the closest Splitter to removed DFF.) An Example is shown in Fig. 8 (a) where the number of added gates in MCID model is 7. Another factor in number of copied gates is reconvergent paths which is dependent on the number of Splitters. So, for the Splitter followed by a removed DFF which cause reconvergent paths with different base distance, the upper bound of number of copied gates can be calculated by summation over the Splitters (e.g., Fig. 8 (d)).

$$\# \text{ of added gates} = \sum_{Spl_i \in Splitter\ cells} 2^{D_{p_{Spl_i}}} - 1 \quad (4)$$

Two examples are shown in Fig. 8 (b) and (c). Fig. 8 (b) shows two independent reconvergent path while the circuit of Fig. 8 (c) has two DFFs in one reconvergent path. Fig. 8 (d) depicts MCID circuit of Fig. 8 (c). As equation (4) shows, the upper bound of added gates are number of Splitters times number of gates before each Splitter (which is, in worst case, a binary tree by height of logical level of Splitter.) Therefore, the MCID model is linearly larger than the original circuit.

We calculate the upper bound for the largest circuit of each type which are 32-bit KSA, 16-bit integer divider and 16-bit array multiplier. We consider them as corner cases and further evaluate the effectiveness of MCID model using them. Table II lists the average verification runtime of ten experiments based on the corner cases, the original number of gates, as well as the upper bound of total number of gates in MCID circuit in different benchmark circuits. The upper bound for MCID circuits was generated according to Section IV.C.2. Results show even for the largest MCID circuits built from benchmark circuits, the proposed LEC still can detect structural errors in a reasonable time.

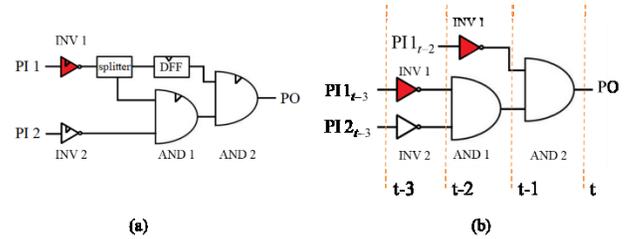

Fig. 7. An example of duplicating a gate within MCID model. (a) SFQ circuit, (b) corresponding MCID model.

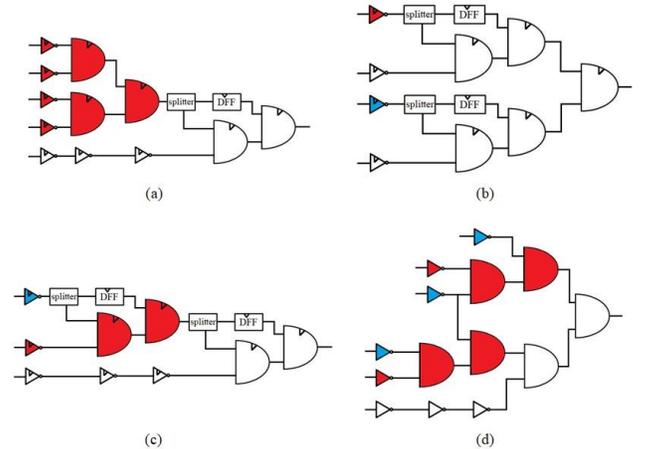

Fig. 8. (a) A 5-level SFQ circuit with one reconvergent path where $D_{p_{Spl}}$ is 4 and the number of added gates in MCID model is 7, (b) a 4-level SFQ circuit with two independent reconvergent path where $D_{p_{Spl}}$ is 2 and the number of added gates is 2 in MCID model which is twice the number of added gates in each reconvergent path, (c) a 5-level SFQ circuit with two DFF in one reconvergent path, (d) Corresponding MCID circuit after removing 2 DFFs. The gates copied in MCID are highlighted by red and blue colors.

*3) Scalability*

In order to show the scalability of the proposed LEC, we compare its runtime to that of ABC which is a scalable verification tool. We used three different types of multipliers including an 8-bit Booth multiplier, an 8-bit Wallace multiplier, and an 8-bit Dadda multiplier as golden models and verified the correct functionality of an 8-bit array multiplier using these golden models. We choose multiplier circuit since it has been shown that checking arithmetic miters is still a challenge in hardware verification [26]. The experimental results are showed in Table III. Table III indicates scalability of our implemented tool as our tool has comparable runtime with respect to ABC. Our results also show that in case of 16-bit multiplier, both ABC and our tool cannot finish verification process in limited time budget, e.g., 12 hours.

By adding the structural error(s), ABC still consider the circuits as equivalent but the proposed LEC can generate the error trace in less than 1 second. The main reason is that ABC only consider the functionality of each gate in its LEC functions. For instance, ABC treats DFFs that are inserted during the synthesis of RSFQ circuits to perform path balancing in beyond-CMOS are considered as simple buffers which is a non-clocked gate.

TABLE II. EFFECTS OF MCID MODEL SIZE ON EFFICIENCY OF THE PROPOSED LEC.

| Benchmark | Type of Error | Run Time | # of Gates | | |
|---|---|---|---|---|---|
| | | | Golden | SFQ | MCID circuit |
| KSA32 | Structural | 0.01 | 449 | 1423 | 1914 |
| ArrMult16 | Structural | 0.12 | 1872 | 5768 | 61409 |
| IntDiv16 | Structural | 0.04 | 1325 | 19261 | 90870 |

TABLE III. RUNTIME OF THE PROPOSED LEC AND ABC.

| Benchmark | Golden Model | LEC Tool | Runtime (s) |
|---|---|---|---|
| ArrMult8 | 8-bit Booth multiplier | proposed LEC | 11.736 |
| | | ABC | 10.413 |
| ArrMult8 | 8-bit Wallace multiplier | proposed LEC | 21.603 |
| | | ABC | 20.611 |
| ArrMult8 | 8-bit Dadda multiplier | proposed LEC | 21.980 |
| | | ABC | 20.568 |

## V. CONCLUSION

We presented a logical equivalence checking framework, based on a novel circuit model called MCID, which is able to perform logic verification of beyond-CMOS technologies independently from the underlying details of signaling and timing. The framework consists of several structural checkers of new technologies constraints satisfaction and build on top of conventional CMOS LEC tools. We additionally presented a multi-cycle input dependency circuit model to explicitly capture the dependence of primary outputs of the circuit on all possible sequences of primary inputs. We further proposed an input timing logic that effectively handled different arrival time of inputs and facilitated the miter circuit generation. We benchmarked the framework on post-synthesis netlists with an RSFQ technology. Results showed a comparative verification time of RSFQ circuit benchmark including ISCAS'85 circuits with respect to ABC tool for similar CMOS circuits. Results confirmed that proposed framework efficiently consider main technological constraints for emerging technologies.

REFERENCES

[1] T. N. Theis and H. S. Philip Wong, "The End of Moore's Law: A New Beginning for Information Technology," *Comput. Sci. Eng.*, vol. 19, no. 2, pp. 41–50, Mar. 2017.

[2] J. A. Hutchby, G. I. Bourianoff, V. V. Zhirnov, and J. E. Brewer, "Extending the road beyond CMOS," *IEEE Circuits Devices Mag.*, vol. 18, no. 2, pp. 28–41, Mar. 2002.

[3] G. Grossing and A. Zeilinger, "Quantum Cellular Automata," *Complex Syst.*, vol. 2, no. 1, pp. 197–208, 1988.

[4] A. Khitun and K. L. Wang, "Non-volatile magnonic logic circuits engineering," *J. Appl. Phys.*, vol. 110, no. 3, p. 034306, 2011.

[5] G. Csaba, A. Imre, G. H. Bernstein, W. Porod, and V. Metlushko, "Nanocomputing by field-coupled nanomagnets," *IEEE Trans. Nanotechnol.*, vol. 1, no. 4, pp. 209–213, Dec. 2002.

[6] R. P. Cowburn, "Room Temperature Magnetic Quantum Cellular Automata," *Science (80-. ).*, vol. 287, no. 5457, pp. 1466–1468, 2002.

[7] K. K. Likharev and V. K. Semenov, "RSFQ Logic/Memory Family: A New Josephson-Junction Technology for Sub-Terahertz-Clock-Frequency Digital Systems," *IEEE Trans. Appl. Supercond.*, vol. 1, no. I, pp. 3–28, 1991.

[8] Y. Harada, H. Nakane, N. Miyamoto, U. Kawabe, E. Goto, and T. Soma, "Basic operations of the quantum flux parametron," *IEEE Trans. Magn.*, vol. 23, no. 5, pp. 3801–3807, 1987.

[9] N. Takeuchi, D. Ozawa, Y. Yamanashi, and N. Yoshikawa, "An adiabatic quantum flux parametron as an ultra-low-power logic device," *Supercond. Sci. Technol.*, vol. 26, no. 3, p. 035010, Mar. 2013.

[10] O. A. Mukhanov, "Energy-Efficient single flux quantum technology," *IEEE Trans. Appl. Supercond.*, vol. 21, no. 3 PART 1, pp. 760–769, Jun. 2011.

[11] Al Geist, "Paving the Roadmap to EXASCALE," *SciDAC Rev.*, vol. 16, pp. 52–59, 2010.

[12] J. Atulasimha and S. Bandyopadhyay, "Bennett clocking of nanomagnetic logic using multiferroic single-domain nanomagnets," *Appl. Phys. Lett.*, vol. 97, no. 17, p. 173105, Oct. 2010.

[13] G. Pasandi and M. Pedram, "PBMap: A Path Balancing Technology Mapping Algorithm for Single Flux Quantum Logic Circuits," *IEEE Trans. Appl. Supercond.*, vol. 29, no. 4, pp. 1–14, Jun. 2019.

[14] G. D. Hachtel and F. Somenzi, *Logic Synthesis and Verification*. Springer Science & Business Media, 2012.

[15] B. L. Synthesis and V. Group, "ABC: A System for Sequential Synthesis and Verification, Release 12/10/06," *Release 70930*, pp. 1–19, 2012.

[16] L. W. Cotten, "Circuit implementation of high-speed pipeline systems," in *Proceedings of the November 30--December 1, 1965, Fall Joint Computer Conference, Part I*, 2008, p. 489.

[17] M. J. Flynn, "Very High-Speed Computing Systems," *Proc. IEEE*, vol. 54, no. 12, pp. 1901–1909, Dec. 1966.

[18] O. Zografos *et al.*, "Wave pipelining for majority-based beyond-CMOS technologies," in *Proceedings of the 2017 Design, Automation and Test in Europe, DATE 2017*, 2017, pp. 1306–1311.

[19] K. Takagi, A. Fujimaki, N. Takagi, Y. Ando, M. Tanaka, and R. Sato, "Design and Demonstration of an 8-bit Bit-Serial RSFQ Microprocessor: CORE e4," *IEEE Trans. Appl. Supercond.*, vol. 26, no. 5, pp. 1–5, 2016.

[20] G. Tang, K. Takata, M. Tanaka, A. Fujimaki, K. Takagi, and N. Takagi, "4-bit Bit-Slice Arithmetic Logic Unit for 32-bit RSFQ Microprocessors," *IEEE Trans. Appl. Supercond.*, vol. 26, no. 1, pp. 1–6, Jan. 2016.

[21] M. D. Ciletti, *Advanced digital design with the Verilog HDL*, vol. 1. Prentice Hall Upper Saddle River, 2003.

[22] G. Pasandi and M. Pedram, "A Graph Partitioning Algorithm with Application in Synthesizing Single Flux Quantum Logic Circuits," *CoRR*, vol. abs/1810.0, 2018.

[23] P. I. Bunyk *et al.*, "High-speed single-flux-quantum circuit using planarized niobium-trilayer Josephson junction technology," *Appl. Phys. Lett.*, vol. 66, no. 5, p. 646, Jan. 1995.

[24] D. S. Holmes, A. L. Ripple, and M. A. Manheimer, "Energy-Efficient Superconducting Computing—Power Budgets and Requirements," *IEEE Trans. Appl. Supercond.*, vol. 23, no. 3, pp. 1701610–1701610, 2013.

[25] N. Katam, S. N. Shahsavani, T.-R. Lin, G. Pasandi, A. Shafaei, and M. Pedram, "Sport lab sfq logic circuit benchmark suite," *Univ. South. California, Los Angeles, CA, USA, Tech. Rep*, 2017.

[26] A. Biere, "Collection of combinational arithmetic miters submitted to the SAT Competition 2016," *SAT Compet.*, vol. 2016, p. 1, 2016.